\documentstyle[11pt,paspconf]{article}

\begin{document}

\title{The Role of Accretion in Forming the Galactic Halo}

\author{Kathryn V. Johnston}
\affil{Institute for Advanced Study, Princeton, NJ 08540}

\begin{abstract}
If the Galaxy formed hierarchically through the accretion of smaller
satellite galaxies we might hope to find signatures of this
in the halo's phase-space distribution.
I review theoretical ideas
about what form these signatures should take,
concentrating in particular on the characteristics of debris from
the tidal disruption of satellites less than
the size of the Large Magellanic Clouds 
and at distances greater than 10 kpc.
Under these circumstances, tidal debris 
can remain aligned in streams of stars close to
the satellite's original orbit
for the lifetime of the Galaxy and we
can understand the evolution of its structure
analytically.
This understanding has many applications to observations and I will
discuss a few examples:  
reconstructing the properties of a long-dead satellite from
its debris; 
measuring the mass loss rate from Galactic satellites through
the density of extra-tidal stars;
and using observations of motions of tidal stream stars
to measure the mass and size of the Milky Way.
\end{abstract}


\section{Introduction}

Theories describing the formation of the Galactic halo can be viewed as 
variations on or combinations of two scenarios.
Based on the kinematics of metal-poor halo field stars
\cite{els} proposed that the halo formed during 
the rapid collapse of an initially uniform primordial density fluctuation.
In contrast, to account for the lack of metallicity gradient
in the outer halo, \cite{sz} proposed that it was formed gradually
through the merging of many sub-galactic sized lumps.
On somewhat larger scales,
fluctuations in the Cosmic Microwave Background radiation and
observations of Large Scale Structure are most closely matched by
simulations of Cold Dark Matter dominated Universes, in which
galaxies form hierarchically from the accretion of smaller satellites.
Observations of individual galactic halos also show substructure that could
be signatures of these disruption events --- both around our own Galaxy
(see \cite{m98} in this volume for a review) and around external galaxies 
(e.g. \cite{s98}).
Hence, current theoretical ideas and observations support a view that
satellite accretion has played some role in the formation of the
Galaxy.

The next decade promises significant progress
in building a global picture of 
halo structure and substructure, with
upcoming satellite missions providing accurate phase-space
coordinates for individual stars in the inner and outer halo.  NASA's Space
Interferometric Mission (SIM), scheduled for 2006, is a pointed
instrument that will detect stars as faint as 20th magnitude with
accuracies of a few $\mu as$.  ESA's Global Astrometric Interferometer
for Astrophysics (GAIA) will survey more than a billion stars across
the entire sky with an astrometric precision of $\le 10\mu as$.  This
represents an improvement over results obtained with the HIPPARCOS
satellite by a factor of about a thousand in accuracy and more than a
million in the volume sampled.
Such a data set should allow us to understand what fraction of the
halo formed in situ and what fraction was  
later added either through minor mergers
(where the mass of the infalling body is a few percent of the mass of the
Milky Way) or satellite accretion (where the satellite has a mass
less than a few tenths of a percent of the mass of the Milky Way).

With sufficient computing power, one can imagine interpreting
substructure in the halo using fully self consistent simulations
of galaxy formation and subsequent merging and accretion events
in a cosmological setting.
Unfortunately, with current hardware and software
it is barely feasible to resolve the mass scales of
even the largest of 
the Milky Way's current population of satellites ($10^6-10^9M_{\odot}$)
with more than a few hundred particles.
Hence, in this review 
rather than trying to place halo formation in a cosmological
context 
I will instead focus on understanding what could be the
observable signatures of the disruption of a small
satellite ($\le 10^{-3} M_{\rm MW}$ where $M_{\rm MW}$ is
the mass of the Milky Way) in the outer halo
($\ge 10$kpc) of our Galaxy.
Although such events may
contribute only a small fraction of the 
Milky Way's total mass, they are nevertheless
interesting because these mass and orbit parameters: 
(i) give timescales for debris dispersal
longer than the age of the Galaxy so
signatures from destruction of such primordial
satellites should still be visible in the halo field star
population; (ii) are comparable to those 
of todays satellites, so we can apply the results to 
features in the halo that could
be associated with the current satellite population;
and (iii) result in the close alignment of 
debris with a single orbit which has interesting implications for the use
of debris to measure the Galactic potential.
For explorations 
of the disruption of small objects in the inner halo see
Helmi \& White (1998) and Harding et al. (1998) 
in this volume, and for a review 
of minor and major mergers
see \cite{b96}.

In \S2 I will discuss how debris from the
disruption of a satellite in orbit around the Milky Way disperses and
in \S3 I will review how this understanding can be applied to interpret
observations of substructure in the halo.
I will summarize these ideas and future prospects in this field in \S4

\section{Debris Dynamics}

\subsection{Simulations} 

\begin{figure}[!htb]
% psfile=#1 vsize=#2 angle=#3 hscale=#4 vscale=#5 hoffset=#6 voffset=#7
\plotfiddle{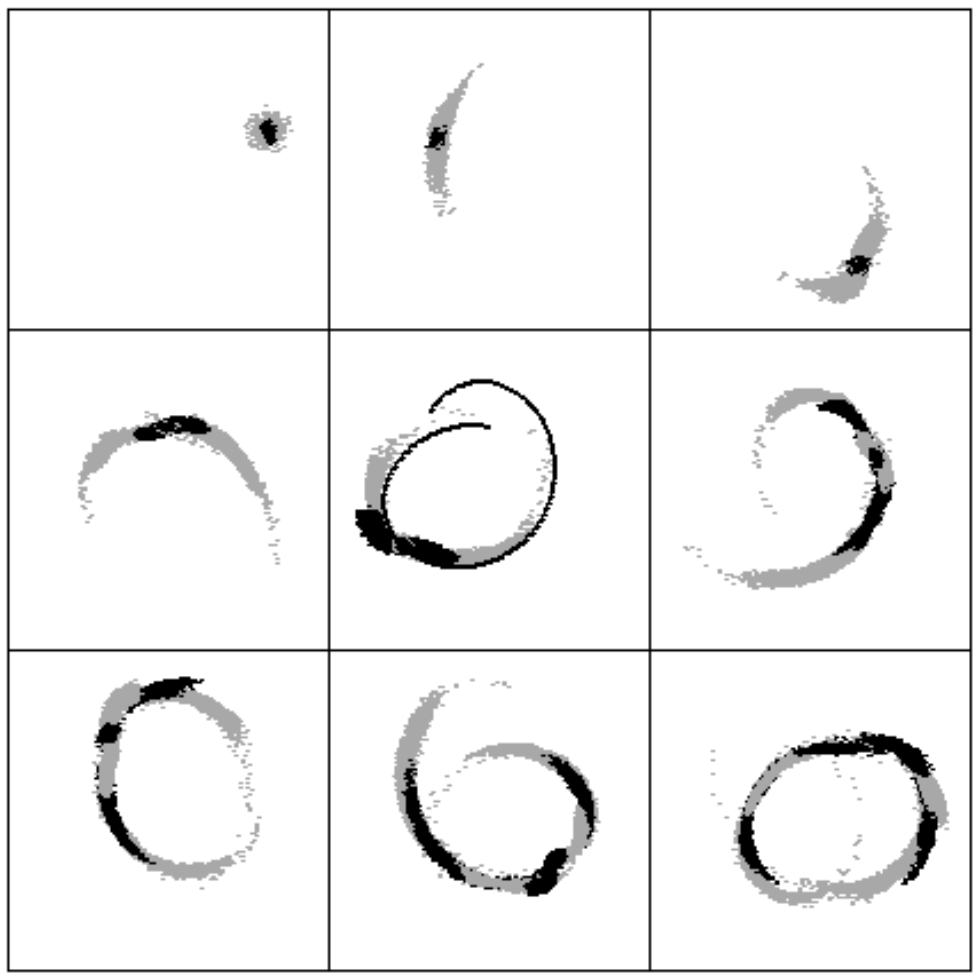}{6.0in}{0}{75}{75}{-230}{-100}
\caption{Evolution of debris from a satellite disrupting while
orbiting around the Milky Way. Each box is centered on
the Galaxy and is 176 kpc on a side. 
The frames are equally spaced over 4.5 Gyrs and
the particles are color-coded with the time
they were torn from the satellite. The orbit is
overlaid as the solid curve in the middle box.  See text for parameters.
}
\label{sim_fig}
\end{figure}

One approach to understanding how debris from satellite disruptions
disperses is to use numerical simulations.
Figure \ref{sim_fig} summarizes the evolution of a satellite in
one such simulation, where the satellite's self-gravity was followed using
a basis-function-expansion code written by \cite{ho92}
and the Milky Way's potential was modeled with bulge, disk and halo
components represented by static, analytic functions (see
\cite{jsh95} for more details of the technique).
The satellite's particles were initially distributed as a Plummer model,
with a mass of $10^8 M_{\odot}$ and scale length of 0.55kpc,
and it was evolved along an orbit that oscillated
between peri- and apo-centric
distances of 30kpc and 55kpc, with a radial time period of 1 Gyr.
The mass and orbit for this satellite were chosen to be
comparable to the dwarf spheroidal satellites we see in orbit
around the Milky Way. However, it's physical
scale was deliberately set so that it 
would lose a significant
amount of mass during course of the simulation.

The panels in Figure \ref{sim_fig} are equally spaced over 4.5 Gyrs.
In the middle panel the orbit of the satellite has been overlaid to emphasize 
that the tidal debris spreads predominantly along the orbit of the
satellite during this time and hence that this alignment should
last for the lifetime of the Galaxy.
Since most of the mass loss from the satellite occurs
at pericenter (where the tidal field of the Milky Way is
strongest), the particles have been coded in greyscale: the light particles
were lost during the first and third pericentric passages, and the dark
particles were lost during the second and fourth.
Note that these sets form distinct populations. The evolution of 
each population 
can be described separately using the formalism outlined in the
next section.

\subsection{Analytic representation of debris dispersal in spherical
potentials}

Tremaine (1993) outlines a simple physical picture of debris dispersal,
which I summarize below. In all equations the first equality
gives the expression for the properties of 
a tidal streamer in a general spherical potential
$\Phi(R)$ and the second equality is for the specific case
of a logarithmic
potential 
\begin{equation}
	\Phi=v_{\rm circ}^2 {\rm{LOG}}(R), 
\label{log}
\end{equation}
as might be appropriate for the halo of
a galaxy with a flat rotation curve
and circular velocity $v_{\rm circ}$.

\begin{figure}[!htb]
\plotfiddle{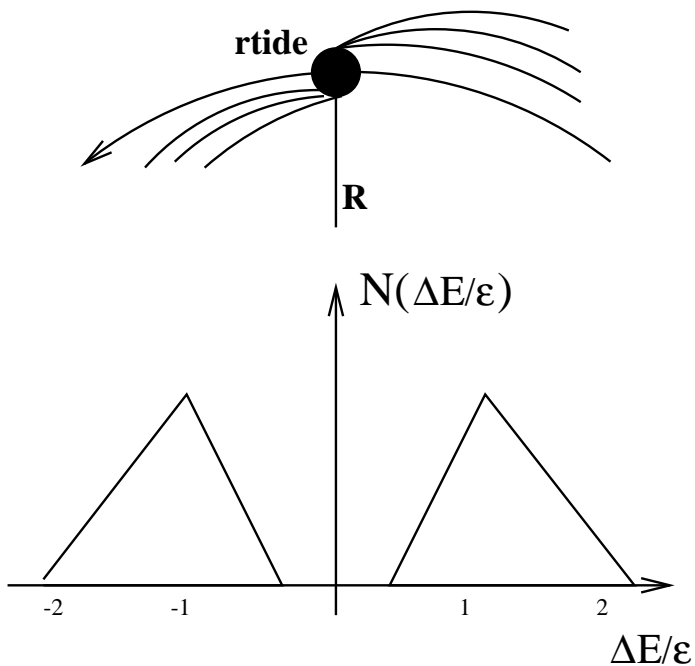}{3.4in}{0}{50}{50}{-120}{0}
\caption{Characteristics of satellite disruption. Upper panel 
shows a cartoon of the spatial distribution and the lower panel is 
the corresponding distribution in scaled energies (see text)}
\label{cartoon_fig}
\end{figure}
Figure \ref{cartoon_fig} is a cartoon representation of the characteristics
seen in the simulation.
The satellite is limited by the Milky Way's tidal field
to a physical scale given by
its {\it tidal radius}
\begin{equation}
\label{rtide}
        r_{\rm tide}=\left({m \over M}\right)^{1/3}R
=\left({Gm \over v_{\rm circ}^2 R}\right)^{1/3}R
\end{equation}
where $m$ is the mass of the satellite system, $R$ is the distance
between the parent and satellite 
and $M$ is the mass of the parent system enclosed within this
radius (\cite{k62}).
The stars escaping from the satellite which lose (gain) energy
move more tightly (less tightly) bound orbits and form
leading (trailing) streams of debris along its orbit.
The energy scale over which these particles are distributed 
is given by
\begin{equation}
\label{epsilon}
        \epsilon=r_{\rm tide}  {d\Phi \over dR}
=\left({Gm \over v_{\rm circ}^2 R}\right)^{1/3} v_{\rm circ}^2.
\end{equation}
This is larger than the internal energy of the satellite by
a factor $(M/m)^{2/3}$ and smaller than the satellite's orbital
energy by a factor $(m/M)^{1/3}$.
When the energies of debris particles  in
simulations of accretion with a wide variety of satellite and
orbit parameters are scaled
by the factor given in equation (\ref{epsilon}) their distributions
are found all found to have the form illustrated in the bottom panel
of Figure \ref{cartoon_fig}.

\begin{figure}[!htb]
\plotfiddle{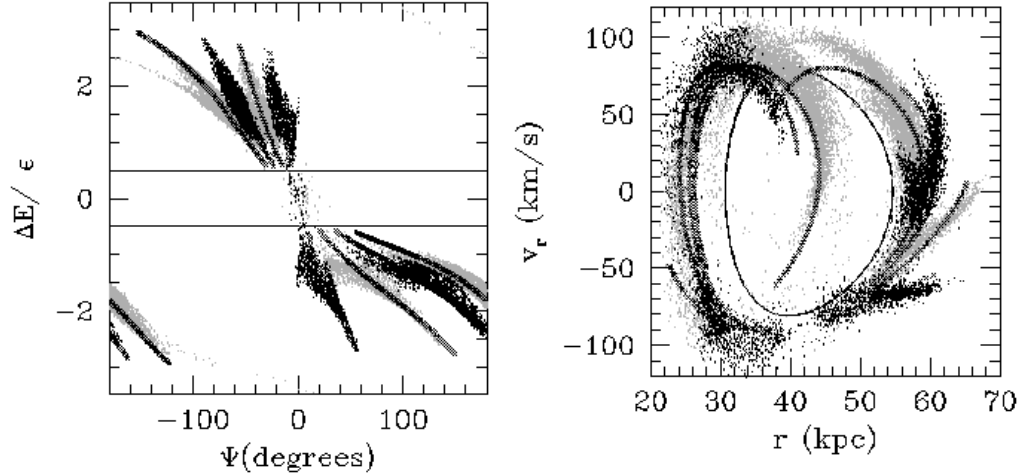}{2.5in}{0}{70}{70}{-210}{-180}
\caption{Left hand panel: energies of debris particles at the end of the 
simulations illustrated in Figure \ref{sim_fig} plotted against their angular
distance from the satellite. Right hand panel: radial velocity of 
debris particles plotted against distance from the parent galaxy. The closed
curve in this panel is the locus of the satellite's orbit.
The solid squares in both panels show the semi-analytic predictions
for the properties of the tidal streamers.}
\label{mod_fig}
\end{figure}
We can approximately predict the spatial distribution of debris lost 
from their energy distribution 
by taking the time period of an orbit perturbed by $\Delta E$ to be
\begin{equation}
	T(E+\Delta E)=\tau(\Delta E)T(E),
\end{equation}
where $T(E)$ is the time period of a circular orbit of energy $E$ and
$\tau$ is a dimensionless function, simply given by 
$\tau(\Delta E)=exp(\Delta E /v_{\rm circ}^2)$ for the logarithmic potential
(i.e. ignoring the weak dependence of time periods on angular momentum).
Then if a satellite on an orbit of energy $E$ loses some mass at time $t=0$
and angular position $\Psi=0$, and moves to position
$\Psi(t)$ at time $t$ later, mass lost with 
energy ($E+\Delta E$) will be at angular position
$\Psi(t/\tau)$.
For full details of this approach, see Johnston (1998).
The success of this simple picture is
illustrated in the left hand panel of
Figure \ref{mod_fig}, which
plots the orbital energies of debris particles relative
to the satellite's orbital energy against angle $\Psi$ along the
orbit at the end of the simulation shown in 
Figure \ref{sim_fig}, with the particles again coded in greyscale with
their escape time. 
Note that few particles have energies $|\Delta E|<\epsilon$
since this is the energy needed to escape from the satellite.
Note also that there is a monotonic trend along each streamer
of particle energy with angle with a small dispersion around it,
confirming that the difference in the orbital energies is primarily
responsible for the debris spreading.
The dark squares overlaid on the particles show the
predictions of the semi-analytic method for the positions
of the streamers in this plane.
The number density along the streamers can be
modeled with similar success by
using the intrinsic distribution of debris in orbital energy
seen in the simulations
(\cite{j98}).

Inspection of the simulations also shows that the full phase-space
position of debris of relative energy $\Delta E$
can be represented by adopting the
velocity of the unperturbed orbit at the same azimuthal phase,
but offsetting the streamer in distance by
\begin{equation}
	\Delta R={\Delta E \over d\Phi/dR}.
\end{equation}
The right-hand panel of Figure \ref{mod_fig}
shows the final positions of particles in the position-radial velocity
plane with the semi-analytic model again overlaid to demonstrate this.

Once the debris spreads to cover 
several oscillations in azimuthal and radial phase the simple approximation
to its phase-space structure outlined above, which is
based only on its energy distribution, will break down.
An exact representation would incorporate all the integrals of motion,
but once the streamers become spatially indistinct 
the value of treating them individually becomes less obvious
(see \cite{hw98} for discussion)

\subsection{Generalizations and Limitations}

The discussion in the previous section was formulated and tested for
spherical satellites with $(m/M)^{1/3} \ll 1$ on bound, non-radial orbits in
the outer halo, which was 
represented by near-spherical, smooth and static potentials. 
In this section I sketch some limits of this approach.

\subsubsection{Small satellite disruption in the inner halo:}
In the simulations discussed so far, 
the dominant source of phase-mixing is that due to
the spreading of debris in one dimension along the orbit itself,
only a few orbital phases are filled during the lifetime of
the Galaxy and the signatures of disruption remain 
distinct in all phase-space dimensions. 
In contrast, in the inner halo,
Helmi \& White (1998) find that debris from satellite 
disruption spreads to cover many orbital phases within a Galactic lifetime,
and that the maximum volume of phase-space filled (in infinite time)
can be found by simply integrating the satellite's orbit.
In addition, they demonstrate that
signatures of disruption can remain
clear in velocity-space even when the streamers are spatially indistinct
(see also \cite{spag98}).

\subsubsection{Satellite geometry:}
If a disk rather than spherical satellite disrupts, similar principles
should apply for the scale at which disruption occurs and for
the way in which debris spreads. However, the distribution in orbital
energies and angular momenta will depend on the internal
phase-space structure of the satellite and this would be reflected in
the details of the density distribution along the tidal streamers.

\subsubsection{Parent galaxy geometry:}
Although the discussion in the previous two sections
used examples of disruption in 
parent galaxy potentials that were dominated by spherical components
I have found similar alignment
of debris with the satellite's orbit in a small number of test
simulations of bound, non-radial orbits in oblate and triaxial potentials.
One can imagine that this alignment would break down if the
debris spread into regions of phase space where the orbit
type abruptly changed (e.g. from a loop to a box orbit)
or became chaotic. In the case of a disk galaxy like
our Milky Way, this is likely to be a concern only in the
inner regions where the potential will be most highly flattened
by the disk
or triaxial due to the influence of the bar.

\subsubsection{Minor mergers and plunging orbits:}
There are many cases of simulations reported
in the literature where the one-dimensional spread of
debris along a single orbit is not seen (e.g.
\cite{hq88}).
In minor mergers or for disruptions 
along radial orbits this is because the condition
$(m/M)^{1/3} \ll 1$ is not satisfied and the debris is released with
a large range in orbital energies, inclinations and phases.
Hence, it quickly spreads to fill two or more spatial
dimensions and cannot be modeled effectively by considering phase-mixing 
along a single orbit.

\subsubsection{Time dependence and smoothness of parent galaxy potential:}
In all of the above cases I have considered
the evolution of a cloud of
test particles phase-mixing along orbits in smooth, static potentials.
This ignores changes in the satellite's orbit due to
dynamical friction, the self-gravity in the debris and
evolution of tidal streamer's due to a lumpy or time-dependent
parent galaxy potential.
The first of these simplifications is justified in the
range of satellite and orbital parameters considered
and the second can be demonstrated to be unimportant (\cite{j98}).
However, the last of these simplifications should
be noted as a concern --- potential fluctuations would affect both
the streamer and satellite orbits. 

\section{Applications}

\subsection{Modeling substructure in our own halo}

\subsubsection{Reconstructing Properties of Long-Dead Satellites Using 
Conserved Quantities}

The longevity of the alignment of tidal streamers
suggests that properties of primordial satellites
could be reconstructed by searching for objects associated with
a single orbit.
This idea was first applied to 
the two-dimensional projected distribution of satellites
by \cite{lb76} and \cite{kd77},
who both noted that several of the dwarf spheroidals 
lie close to the Great Circle defined by the Large Magellanic Clouds
and the Magellanic Stream.
Later studies suggested possible associations 
of globular clusters with each other and with the larger
satellites (\cite{lr92,m94,fp95,lb2}) and
\cite{jhb96} applied the same principles to
develop a general algorithm
that searches for Great Circle alignments in stellar data sets.
\cite{it98} have recently found
perhaps the most compelling example of satellite debris seen
in projection using this latter method ---
a set of carbon stars aligned with the direction of elongation
of the Sagittarius dwarf galaxy.

The two-dimensional approach implicitly assumes that the halo
potential is only mildly oblate and that debris will lie
close to a plane (or Great Circle in projection on the
sky).
Lynden-Bell \& Lynden-Bell (1995) show how, with the assumption of a specific
form for the Galactic potential,
the addition of distance and radial velocity measurements
can provide further tests of object association by
requiring that they have similar energy and angular momentum.
The accurate proper motion measurements 
with the upcoming SIM and GAIA satellite missions (see \S 1)
will allow the measurement of angular momenta directly in the inner halo, 
which can alone serve as a fair discriminant of streamers
from the field population (see \cite{hzd98}).

\subsubsection{Finding Local Features}

\begin{figure}[!htb]
\plotfiddle{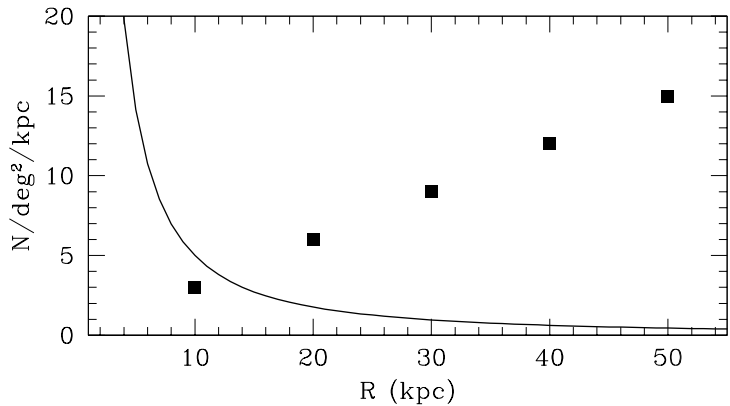}{2.0in}{0}{50}{50}{-180}{-200}
\caption{The solid line shows the number of giant branch stars
per kpc along the line of sight in a degree-squared patch of sky
for a smooth halo distribution.
Each solid square shows the equivalent density of debris left from the
disruption of a satellite at that distance 10 Gyrs ago.}
\label{dens_fig}
\end{figure}

The methods in the previous section all used conserved quantities
to discriminate likely members of streamers over large
areas of the sky.
On smaller (square-degree)
scales, phase-space substructures in the form of moving
groups of stars in the inner halo have already been detected (see
\cite{mhm96} for a review and \cite{hw98} for interpretation).
The first results from two surveys to identify and measure distances to
and radial velocities of giant stars to far greater depths in
the halo also show significant substructure (\cite{m98,spag98}).
Figure \ref{dens_fig} illustrates the power of such studies. The
solid curve shows an estimate for the number of giant stars
per square degree per kpc along the line of sight
if the outer halo followed the smooth $\rho \propto r^{-3.5}$
density profile measured in the inner halo
(\cite{f96}).
Each of the solid squares show the density of debris
from a satellite disrupted 10 Gyrs ago at the given distance
from the Galactic center, estimated using the scalings
given in \S 2.2.
The figure suggests that stellar debris from such an event
should be easily identified above the 
smooth background with the addition of distance measurements
alone.

\subsection{Measuring Mass Loss Rates from Galactic Satellites}

\begin{figure}[!htb]
% psfile=#1 vsize=#2 angle=#3 hscale=#4 vscale=#5 hoffset=#6 voffset=#7
%\plotfiddle{johnston_fig5.eps}{2.in}{0}{45}{45}{-150}{-80}
\plotfiddle{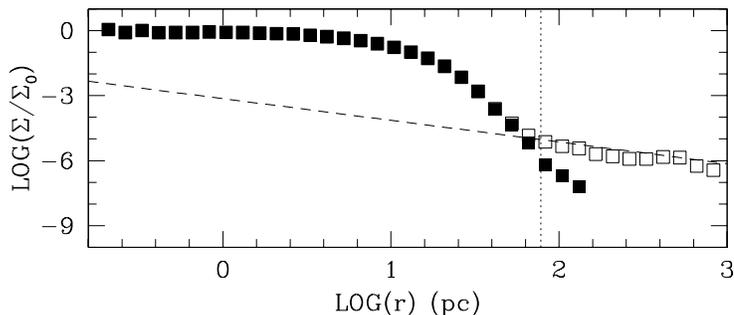}{1.8in}{0}{50}{50}{-180}{-100}
\caption{Annularly averaged number surface density
from a simulation of satellite disruption ``observed'' 
from the viewpoint of the center of the Galaxy after several Gyrs
The closed (open) symbols show the profile recovered if only bound (all)
stars are considered. The dashed line shows the density of
``extra-tidal'' stars (i.e. open symbols) predicted from our understanding of
debris dynamics}
\label{mloss_fig}
\end{figure}

Number count profiles of many Galactic and some extra-galactic 
satellite systems show evidence for associated stars beyond
the cut-off in density that is identified as the point of
tidal limitation (e.g. Irwin \& Hatzidimitriou 1995,
Grillmair et al. 1995). 
These ``extra-tidal'' stars are assumed to be debris lost from
the satellite due to heating or stripping by the Galactic tidal
field or (in the case of globular clusters) evaporation
of stars over the tidal boundary.
Figure \ref{mloss_fig} shows the annularly averaged number surface density
from a simulation of satellite disruption ``observed'' 
from the viewpoint of the center of the Galaxy after several Gyrs.
The closed (open) symbols show the profile recovered if only bound (all)
stars are considered.
Clearly there is a break in the open symbols at the radius 
where the analysis becomes dominated by unbound stars (approximately
indicated by the vertical dotted line) confirming 
the interpretation of ``extra-tidal'' stars as debris.
\cite{jsh98} apply the same principles outlined in \S2.2
to estimate the expected density of the tidal debris beyond the
tidal radius from the mass-loss rate seen in the simulation and
this estimate is overlaid in dashed lines on the profile shown in
Figure \ref{mloss_fig}.
This method reproduces the extra-tidal features in
all simulations tested with similar success.
This suggests that our understanding of debris dynamics can be used to
measure the mass loss rate from a Galactic satellite using the 
observed density of extra-tidal stars.
When tested on the simulations, this approach recovered the
mass loss rate to within a factor of two.
When applied to observations it can be used to provide
constraints on dynamical models of individual satellites and
to directly measure the current destruction rate of the Galactic
satellite system.

\subsection{Measuring Potentials}

The alignment of tidal streams along a single orbit make
them uniquely useful as probes of the Galactic potential,
an idea thast has been
noted by a number of authors (\cite{lb82}; \cite{ksh96};
\cite{g98}).  By assuming that several of the dwarf spheroidal
satellites are tidal debris, Lynden-Bell (1982) was able to obtain an
estimate of the mass of the Milky Way.  Similarly, if the Magellanic
Stream consists of gas tidally stripped from the Large and Small
Magellanic Clouds, it can be used to constrain the potential
(\cite{mf80}; \cite{llb82}).  While these results are not definitive
measurements owing to the controversial nature of the Magellanic
Stream (\cite{md94}) and association
of the dwarf spheroidals, they demonstrate the power of this approach.

The growing observational evidence for the existence of stellar tidal
streamers in the halo (\cite{m98}), coupled with the 
prospect of highly accurate
measurements of stellar proper motions in the outer halo
using the SIM and GAIA satellite missions to be launched 
within the next decade
motivates a re-examination of this idea.
Assuming measurements of just 100 stars in a tidal streamer associated with
a known Galactic satellite could be made with these
upcoming missions, \cite{jzsh98} found that 
the circular velocity and axis ratios of the halo could be recovered
to within a few percent - an order of magnitude improvement
over previous methods (see \cite{zjsh98} in this
volume for a summary of the method).

\section{Summary}

In this review I have looked at the characteristics of
signatures we would expect in
the outer halo's phase-space distribution from the accretion of
small satellites. Under these circumstances, tidal streamers from
the disruption of the satellite are long-lived, the rate of debris
dispersal is well-understood and their structure
can be modeled using simple physical arguments.
We can
use the properties of any streamers observed in the halo 
to reconstruct the characteristics of the
primordial object from which they came.
If they are associated with a particular satellite, we
can use their density to measure the rate at which the satellite 
is currently disrupting.
Finally, even if the halo is dominated by a smooth component and accretion
of small satellites is unimportant, a small number stars in tidal streamers
can be used to measure the Galactic potential with far greater accuracy
than the same number in the random population.

The next decade promises to be an exciting one for progress in this
field. Preliminary results from current observational surveys
hint that the outer halo is indeed far from smooth, and
future satellite missions promise to extend these local samples to 
build a truly global picture of Milky Way structure and substructure,
with accurate measurements of all phase-space coordinates in the
inner halo and five of the six phase-space dimensions further
out (the error in the measurement of the distance $D$ to an object
is $\sim(D/20kpc)^2$).
With the prospect of applying their results to
these observational data sets,
theorists can look beyond the classic static
models of Milky Way structure, and beyond
the simple picture of debris dispersal from small satellites
I have outlined in this review and hope to constrain a
fully self-consistent picture of Galaxy formation and evolution.

\acknowledgments
I thank Scott Tremaine and Amina Helmi for helpful comments on this
paper and the conference organisers both for the invitation to and
accommodation at the conference.
This work was supported by funds from the
Institute for Advanced Study.

\end{document}